\documentstyle[12pt]{article}

\def\be{\begin{equation}}
\def\ee{\end{equation}}
\def\bea{\begin{eqnarray}}
\def\eea{\end{eqnarray}}

\newfont{\frak}{eufm10 scaled\magstep1}
\newcommand{\fra}[1]{\mbox{\frak #1}}
\newfont{\fraksection}{eufm10 scaled\magstep2}
\newcommand{\frasection}[1]{\mbox{\fraksection #1}}

\def\F{{Freudenthal}}

\def\su{{\fra{su}}}
\def\sl{{\fra{sl}}}
\def\sy{{\fra{sy}}}
\def\my{{\fra{my}}}
\def\so{{\fra{so}}}
\def\sp{{\fra{sp}}}
\def\sa{{\fra{sa}}}

\def\saly{{\frasection{sa, sl, sy}}}

\font\black=msbm10 scaled\magstep1
\def\extra #1{\hbox{{\black #1}}}

\def\R{{\hbox{{\extra R}}}} 
\def\C{{\hbox{{\extra C}}}} 
\def\Q{{\hbox{{\extra H}}}} 
\def\OO{{\hbox{{\extra O}}}} 
\def\K{{\hbox{{\extra K}}}} 

\def\I{{\hbox{{\extra I}}}} 

\def\X{{\hbox{{\extra X}}}} 
\def\J{{\hbox{{\extra J}}}} 
\def\M{{\hbox{{\extra M}}}} 
\def\H{{\hbox{{\extra H}}}} 
\def\E{{\hbox{{\extra E}}}} 

\def\diag{\,\mbox{diag}\,}
\def\mytimes#1{#1\!\times\!#1}

\def\w{\omega}
\def\wi{\w_1}
\def\wii{\w_1\w_2}
\def\win{\w_1\dots\w_N}

\def\npo{N\!+\!1}

\def\hquadback{\!\!} 
\def\tquadback{\!\!\!}

\begin{document}

\thispagestyle{empty}

 \hfill \  
\bigskip\bigskip

\begin{center}

{\LARGE{\bf{``Cayley-Klein" schemes for}}} 

 {\LARGE{\bf{real Lie algebras and}}} 

{\LARGE{\bf{Freudhental Magic Squares}}}

\end{center}

\bigskip\bigskip

\begin{center}
Mariano Santander 

{ \em {   Departamento de F{\'{\i}}sica Te\'orica, Facultad de Ciencias,
}\\ 
 Universidad de Valladolid,  E-47011, Valladolid, Spain \\ 
email: santander@cpd.uva.es}
\end{center}

\begin{center}
Francisco J. Herranz  

{\em { Departamento de F{\'{\i}}sica, E. U. Polit\'ecnica, }
\\  Universidad de Burgos, E-09006, Burgos, Spain \\ 
email: fteorica@cpd.uva.es}
 \end{center}

\bigskip\bigskip

\begin{abstract}
We introduce three ``Cayley-Klein" families of Lie algebras through
realizations in terms of either real, complex or quaternionic matrices.
Each family includes simple as well as some limiting quasi-simple real
Lie algebras. Their relationships  naturally lead to an infinite family
of $\mytimes 3$ {\F}-like magic squares, which relate algebras in the
three CK families. In the lowest dimensional cases suitable extensions
involving octonions are possible, and for $N=1, 2$, the ``classical"
$\mytimes 3$ {\F}-like squares admit a $\mytimes 4$ extension, which
gives the original \F\ square and the Sudbery square. 
\end{abstract}

\newpage


\section{The $\saly$ CK families}

Consider the {\em real} matrices of order $\npo$ given by:
\be
{J_{ab}\!=\!\left( \hquadback \begin{array}{ccccc}  
          & \cdot &          & \!\cdot         &          \\[-0.2cm]
\cdot\!\! & \cdot &\!\!\cdot & \!\!- \w_{ab}   &\!\!\cdot \\[-0.2cm]
          & \cdot &          & \!\cdot         &          \\[-0.15cm]
\cdot\!\! &   1   &\!\!\cdot & \!\!\!\cdot\ \cdot \ \cdot
                                               &\!\!\cdot \\[-0.15cm]
           &\cdot &           &\!\cdot         &         
\end{array} \hquadback \right)\!\!,}
\ \ 
{M_{ab}\!=\!\left( \hquadback \begin{array}{ccccc}  
          & \cdot &           &\!\cdot              &          \\[-0.2cm]
\cdot\!\! & \cdot &\!\!\cdot  &\!\!\!\cdot\ \w_{ab} &\!\!\cdot \\[-0.2cm]
          & \cdot &           &\!\cdot              &          \\[-0.15cm]
\cdot\!\! &   1   &\!\!\cdot  & \!\!\!\cdot\ \cdot \ \cdot 
                                                    &\!\!\cdot\\[-0.15cm]
           &\cdot &           &\!\cdot              &         
\end{array} \hquadback \right)\!\!,}
\ \ 
{H_m\!=\!\left( \hquadback \begin{array}{cccc}
    -1   &                & \hquadback \cdot  &       \\[-0.2cm]
         &                & \hquadback \cdot  &       \\[-0.15cm]
\cdot    &\tquadback\cdot & \hquadback 1      & \cdot \\[-0.2cm]
         &                & \hquadback \cdot  &         
\end{array} \hquadback \right)\!\!,} 
\ \ 
{E_0\!=\!\left( \hquadback \begin{array}{cccc}
   1  &       &        &     \\[-0.2cm]
      &       &        &     \\[-0.15cm]
      &       &        &     \\[-0.2cm]
      &       &        &         
\end{array} \hquadback\hquadback \right)} 
\label{jmhe}
\ee

\noindent where $a, b= 0, 1, \dots, N, \ a<b, \ m=1, \dots, N$, the matrix
indices range from $0, 1, \dots, N$, the dotted rows and columms are
those with row or column indices $a, b$ or $m$, and 
$\w_{ab}:=\w_{a+1}\w_{a+2}\cdots\w_b$ depend on $N$ non-zero real
coefficients 
$\w_1,\dots,\w_N$. 

We apply two basic procedures to these matrices to  build up a set of
matrices over $\R,\C, \Q$ which contain sets of generators for {\em all}
simple classical real Lie algebras: \ 

\noindent $\bullet$ Let $X$ be a real matrix in (\ref{jmhe}). In the
complex case, let $X^1:=iX$. In the quaternion case, let $X^1:=iX, \
X^2:=jX, \ ,X^3:=kX$. Generically, these new matrices will be denoted
$X^\alpha$, where the range of
$\alpha$ is none for $\R$, $1$ for $\C$ and $1,2,3$ for $\Q$. 

\noindent $\bullet$ For any $X$ in the previous list,  define matrices
$\X, \X_{;\lambda}, \ \lambda=1,2,3$ of order $2(\npo)$:
\be
\X=\left( \!\begin{array}{cc} X & \\ & \!X \end{array} \!\right), 
\quad 
\X_{;1}=\left( \!\begin{array}{cc} & \!X\\ -X &  \end{array} \!\right), 
\quad 
\X_{;2}=\left( \!\begin{array}{cc} & \!X\\ X &  \end{array} \!\right), 
\quad 
\X_{;3}=\left( \!\begin{array}{cc} X & \\ & \!\!-X \end{array} \!\!\right)
\ee

Let $A$ be any matrix of order $r$ over either $\K=\R, \C, \Q$, and let
$G$ denote the symmetric or antisymmetric real matrix of an hermitian or
skew-hermitian product in the space $\K^r$. The matrix
$A$ will be called $G$-antihermitian if $A^\dagger G + G A^\dagger =0$,
and $G$-hermitian if $A^\dagger G - G A^\dagger =0$. With the choices 
$I_\w = \diag(1, \w_{01}, \w_{02}, \dots, \w_{0N})$ and
$\I_\w=\I_{;1}$, the matrices $J_{ab}, M_{ab}^\alpha, H_m^\alpha,
E_0^\alpha$ are $I_\w$-antihermitian, and 
$ \J_{ab}$, $\M_{ab;\lambda}$, $\H_{m;\lambda}$, $\E_{0;\lambda}$, 
$\J_{ab;\lambda}^\alpha$, $\M_{ab}^\alpha$, $\H_{m}^\alpha$,
$\E_{0}^\alpha $ are $\I_\w$-antihermitian, no matter of whether 
$\w_i=0$ or not.

Now we define the three ``classical" CK series of algebras as follows
\cite{tesis}:

$\bullet$ $\sa_{\win}(\npo, \K)$, the {\em special antihermitian} CK
algebra over $\K$ is the quotient of the Lie algebras of $\mytimes{\npo}$
$I_\w$-antihermitian matrices over $\K$ by its center. They can be
realized as the Lie algebra of all
$I_\w$-antihermitian matrices over $\K$ if $\K=\R, \Q$ and as the
subalgebra determined by the condition $\hbox{tr}X= 0$ if $\K=\C$. 

$\bullet$ $\sl_{\win}(\npo, \K)$, the {\em special linear} CK algebra
over $\K$ is the quotient of the Lie algebra of all $\mytimes{\npo}$
matrices over $\K$ by its center. It can be realized as the Lie
subalgebra of the Lie algebra of all $\mytimes{\npo}$ matrices over $\K$
determined by the condition $\hbox{tr}X= 0$ if $\K=\R, \C$ and by the
condition $\hbox{Re(tr}X)= 0$  when $\K = \Q$. 

$\bullet$ $\sy_{\win}(2(\npo), \K)$, the {\em special symplectic
antihermitian} CK algebra over $\K$ is the quotient of the Lie algebra of
$\I_\w$-antihermitian matrices of order
$2(\npo)$ over $\K$ by its center. It is the analogous to the first
family  when the metric matrix is the antisymmetric
$\I_\w$. It can be realized again as the Lie algebra of all
$\I_\w$-antihermitian matrices if $\K=\R, \Q$ and as the subalgebra with
$\hbox{tr}\X= 0$ if $\K=\C$. 

The notation $\sa$ has been  used in \cite{Sud} and the notation $\sy$ is
new, although of course the algebras so denoted are not. The CK algebras
with $\win={+\dots+}$ are isomorphic to the ones usually denoted by the
same symbol and without any $\w$ subscript. The translation to the
standard notation is as follows: 
\be
\begin{tabular}{llll}
$ \qquad $ & 
$ \K=\R $ &
$ \K=\C $ &
$ \K=\Q $ \\
$ \sa_{+\dots+}(\npo, \K) \qquad $ &
$ \so(\npo) $ & 
$ \su(\npo) $ &
$ \sp(\npo) $ \\
$ \sl_{+\dots+}(\npo, \K) \qquad $ &
$ \sl(\npo, \R) $ & 
$ \sl(\npo, \C)$ & 
$ \su^*(2(\npo)) $ \\
$ \sy_{+\dots+}(2(\npo), \K) \qquad $ &
$ \sp(2(\npo), \R) $ & 
$ \su(\npo,\npo) $ & 
$ \so^*(4(\npo)) $ \\
\end{tabular}
\ee

When the $\w_i$ are not all positive, the CK algebras in the three $\sa$
series are isomorphic to the non-compact real forms
$\so(p,q)$, $\su(p,q)$, and $\sp(p,q)$. When some $\w_i=0$, the general CK
algebras $\sa_{\win}(\npo, \K)$, etc. are defined in such a way that each
$\w_i =0$ corresponds to a contraction; lack of space precludes giving
details.

The CK Lie algebras $\sa, \sl, \sy$ over the three associative division
algebras $\R, \C, \Q$ can be generated by means of adequate choices of
matrices, as given in the Table:

\medskip

{\noindent 
\begin{tabular}{llll}
\hline
Lie algebra & 
\multicolumn{3}{l}{has a linear basis given by the matrices} \\
\hline 
 & \K=\R & \K=\C & \K=\Q \\
$ \sa_{\win}(\npo, \K)$ & 
$ J_{ab}$ & 
$ J_{ab}, M_{ab}^1, H_m^1$ & 
$ J_{ab}, M_{ab}^\alpha, H_m^\alpha, E_0^\alpha$ \\ 
$ \sl_{\win}(\npo, \K)$ & 
$ J_{ab}, M_{ab}, H_m $ & 
$ J_{ab}, M_{ab}, H_m, $ & 
$ J_{ab}, M_{ab}, H_m, $ \\ 
\ &
$ \ $ &
$ \quad J_{ab}^1, M_{ab}^1, H_m^1 $ &
$ \quad J_{ab}^\alpha, M_{ab}^\alpha, H_m^\alpha, E_0^\alpha$  \\
$ \sy_{\win}(2(\npo), \K)\!\!\! $ & 
$ \J_{ab}, \M_{ab ;\lambda}, \H_{m ;\lambda}, \!\! $ & 
$ \J_{ab}, \M_{ab ;\lambda}, \H_{m ;\lambda}, \E_{0 ;\lambda}, \!\! $ & 
$ \J_{ab}, \M_{ab ;\lambda}, \H_{m ;\lambda}, \E_{0 ;\lambda}, \!\! $ \\
$  \   $ & 
$ \qquad\qquad\quad \E_{0 ;\lambda} $ & 
$ \quad \J_{ab ;\lambda}^1, \M_{ab}^1, \H_{m}^1 \!\! $ & 
$ \quad \J_{ab ;\lambda}^\alpha, \M_{ab}^\alpha, \H_{ma}^\alpha,
                                           \E_{0}^\alpha  \!\! $\\
\hline
\end{tabular}}
 
\medskip


\section{The ``classical" $(\npo)$-d {\F}-like square }

The former table looks simpler if each Lie algebra is given as the Lie
span (instead of the linear span) of as few elements as possible. A
minimal choice is: 
 
\medskip

{\noindent 
\begin{tabular}{llll}
\hline
${}\!\!$Lie algebra & 
\multicolumn{3}{l}{is the Lie span of the generators} \\
\hline 
 & \K=\R & \K=\C & \K=\Q \\
$ \!\!\sa_{\win}(\npo, \K)$ & 
$ J_{ab}$ & 
$ J_{ab}, M_{ab}^1$ & 
$ J_{ab}, M_{ab}^1, M_{ab}^2$ \\ 
$ \!\!\sl_{\win}(\npo, \K)$ & 
$ J_{ab}, M_{ab} $ & 
$ J_{ab}, M_{ab}, M_{ab}^1  $ & 
$ J_{ab}, M_{ab}, M_{ab}^1, M_{ab}^2  $ \\ 
$ \!\!\sy_{\win}(2(\npo), \K)$ & 
$ \J_{ab},\! \M_{ab ;2},\! \M_{ab ;1} \!\! $ & 
$ \J_{ab},\! \M_{ab ;2},\! \M_{ab ;1},\! \M_{ab}^1 \!\! $ & 
$ \J_{ab},\! \M_{ab ;2},\! \M_{ab ;1},\! \M_{ab}^1,
                                      \! \M_{ab}^2 \!\! $ \\[0.2cm] 
\hline 
\end{tabular}}

\medskip \medskip

This Table shows a rather unexpected and remarkable symmetry between rows
and columns. Since $\R \subset \C \subset \Q$, each algebra is in the
obvious way a subalgebra of those at its left. And each algebra is also a
subalgebra of those below it, provided we have made the isomorphic
identifications 
$J_{ab} \to \J_{ab}, M_{ab} \to \M_{ab\,;2},  
M_{ab}^1 \to \M_{ab}^1, M_{ab}^2 \to \M_{ab}^2$. 
This is required since $\sy$ is a group of matrices of dimension twice
that of $\sl$.   
 
We shall better observe this symmetry if we move from the left to the
right and from the top to the bottom. We realize that in each step new
generators appears. Let us illustrate this idea as follows. As we move
from the top $\sa$ to the bottom $\sy$, $M_{ab}$ appears in the first
step ($\sa \to \sl$) and $\M_{ab\,;1}$ in the second ($\sl \to \sy$). 
This behaviour is the {\em same} for the three columns $\R, \C, \Q$. On
the other hand, moving from left to right, in the transition  $\R \to \C$
we always add $M_{ab}^1$, and in the transition $\C \to \Q$ we add
$M_{ab}^2$. This behaviour appears in the {\em three} rows. 

This symmetry suggests to consider a $\mytimes 3$ square of CK algebras,
with rows labeled by $\R, \C, \Q$ and columns by $\sa, \sl, \sy$. Each
site in the Table below contains the generic algebra in the CK family, the
complete list of their basis generators, and the Cartan class of the
corresponding simple Lie algebras. 

\medskip

{\noindent\hfill
\begin{tabular}{lll}
\hline
$ B_{N/2} $ or $ D_{(\npo)/2}$  & 
$ A_N $                    & 
$ C_{\npo}$                 \\
$ \fra{sa}_{\win}(\npo, \R) $ & 
$ \fra{sa}_{\win}(\npo, \C) $ & 
$ \fra{sa}_{\win}(\npo, \Q) $ \\
$ J_{ab}$ & 
$ J_{ab}, M_{ab}^1, H_m^1$ & 
$ J_{ab}, M_{ab}^\alpha, H_m^\alpha, E_0^\alpha$ \\ 
\hline
$ A_N $                      & 
$ A_N\oplus A_N $            & 
$ A_{2(\npo)-1} $             \\
$ \sl_{\win}(\npo, \R)$ & 
$ \sl_{\win}(\npo, \C)$ & 
$ \sl_{\win}(\npo, \Q)$ \\
$ J_{ab}, M_{ab}, H_m $ & 
$ J_{ab}, M_{ab}, H_m, $ & 
$ J_{ab}, M_{ab}, H_m, $ \\ 
$ \quad $ &
$ \quad  J_{ab}^1, M_{ab}^1, H_m^1 $ &
$ \quad  J_{ab}^\alpha, M_{ab}^\alpha, H_m^\alpha, E_0^\alpha$  \\
\hline
$ C_{\npo} $                      & 
$ A_{2(\npo)-1} $                 & 
$ D_{2(\npo)}   $             \\
$ \sy_{\win}(2(\npo), \R)$ & 
$ \sy_{\win}(2(\npo), \C)$ & 
$ \sy_{\win}(2(\npo), \Q)$ \\
$ \J_{ab}, \M_{ab \,;\lambda}, \H_{m \,;\lambda}, \E_{0 \,;\lambda} $ & 
$ \J_{ab}, \M_{ab \,;\lambda}, \H_{m \,;\lambda}, \E_{0 \,;\lambda}, $ & 
$ \J_{ab}, \M_{ab \,;\lambda}, \H_{m \,;\lambda}, \E_{0 \,;\lambda}, $ \\
$ \quad  $ & 
$ \quad \J_{ab;\lambda}^1, \M_{ab}^1, \H_{m}^1 $ & 
$ \quad \J_{ab;\lambda}^\alpha, \M_{ab}^\alpha, 
                   \H_{m}^\alpha, \E_{0}^\alpha $ \\
\hline
\end{tabular}\hfill}

\medskip

Dimension checking is easy. Each $J_{ab}$ or $M_{ab}$ counts as
$N(N+1)/2$, each $H_{m}$ as $N$, and each $E_0$ as one. If the three
columns are labeled by $p=1,2,4$ and the three rows by $q=1,2,4$, the
dimension of the algebra at site $p,q$ is 
\be
\hbox{dim}(p,q) = pq \frac{N (N+1)}{2} + (p+q-2) N + (0,0,3)_p + (0,0,3)_q
\ee
where the symbols $(0,0,3)_p$ or $(0,0,3)_q$ refers to the 3 extra
generators $(E_0^\alpha)$ or $(\E_{0\,;\lambda})$ appearing respectively
when $\K=\Q$ or in the $\sy$ case. There are similar expressions for the
characters of the CK algebras involved. We emphasize that for each choice
of values $\win$, the algebras included in the square are different.
Finally, when some $\w_i=0$ then the algebras are not simple; nevertheless
the properties of the square are maintained in all cases, so this
($\win$)-square includes a ``compact", several ``non-compact" and
additional ``non-simple" versions. 


\section{The lowest-dimensional ``classical" \F\ squares 
         and their extensions to include exceptional algebras.}

The first two squares, $N=1,2$ allow an extension by introducing an
additional $p=8$ column, associated to octonions $\OO$, and an additional
$q=8$ row with the so-called metasymplectic algebras, denoted here $\my$.
Lie  algebras worth of the names 
$\sa(\npo, \OO), \sl(\npo, \OO), \sy(2(\npo), \OO), \my(2(\npo), \OO)$ 
are only possible when $N=1, 2$ and due to the nonassociativity of $\OO$
its proper definition requires some  approach alternative to the one
sketched in Sec. 1 (for $\sa$ and $\sl$ this is done in \cite{Sud}). 

In the $N=2$ case, we get the original \F\ square \cite{FT}, which has
several versions most of which are obtained by particular choices of the
constants $\w_1, \w_2$ in: 

\medskip

{\noindent\hfill 
\begin{tabular}{llll}
\hline
$ B_1 \! \equiv \! A_1 \! \equiv \! C_1 \!\equiv \!E_1$  & 
$ A_2$  & 
$ C_3 \! \equiv \! B_2 $  &
$ F_4 $  \\
$ \fra{sa}_{\wii}(3, \R) $ & 
$ \fra{sa}_{\wii}(3, \C) $ & 
$ \fra{sa}_{\wii}(3, \Q) $ & 
$ \fra{sa}_{\wii}(3, \OO) $ \\
\hline
$ A_2$ & 
$ A_2 \oplus A_2 $  & 
$ A_5 $  &
$ E_6 $  \\
$ \fra{sl}_{\wii}(3, \R) $ & 
$ \fra{sl}_{\wii}(3, \C) $ & 
$ \fra{sl}_{\wii}(3, \Q) $ & 
$ \fra{sl}_{\wii}(3, \OO) $ \\
\hline
$ C_3 \! \equiv \! B_2 $  & 
$ A_5 $  & 
$ D_6 $  &
$ E_7 $  \\
$ \fra{sy}_{\wii}(6, \R) $ & 
$ \fra{sy}_{\wii}(6, \C) $ & 
$ \fra{sy}_{\wii}(6, \Q) $ & 
$ \fra{sy}_{\wii}(6, \OO) $ \\
\hline
$ F_4 $  & 
$ E_6 $  & 
$ E_7 $  &
$ E_8 $  \\
$ \fra{my}_{\wii}(6, \R) \qquad $ & 
$ \fra{my}_{\wii}(6, \C) \qquad $ & 
$ \fra{my}_{\wii}(6, \Q) \qquad $ & 
$ \fra{my}_{\wii}(6, \OO) $ \\
\hline
\end{tabular}\hfill}

\medskip

Notice that reflection in the main diagonal corresponds to a change of
real form. When $N=1$, the extended ``classical" {\F}-like square is:

\medskip

{\noindent\hfill 
\begin{tabular}{llll}
\hline
$ D_1 $  & 
$ A_1 \! \equiv \! B_1 \!\equiv \!C_1 \!\equiv \!E_1$  & 
$ C_2 \! \equiv \! B_2 $  &
$ B_4 $  \\
$ \fra{sa}_{\wi}(2, \R) $ & 
$ \fra{sa}_{\wi}(2, \C) $ & 
$ \fra{sa}_{\wi}(2, \Q) $ & 
$ \fra{sa}_{\wi}(2, \OO) $ \\
\hline
$ A_1 \! \equiv \! B_1 \!\equiv \!C_1 \!\equiv\! E_1$ & 
$ A_1 \oplus A_1 \! \equiv \! E_2 $  & 
$ A_3 \! \equiv \! D_3 $  &
$ D_5 \! \equiv \! E_5 $  \\
$ \fra{sl}_{\wi}(2, \R) $ & 
$ \fra{sl}_{\wi}(2, \C) $ & 
$ \fra{sl}_{\wi}(2, \Q) $ & 
$ \fra{sl}_{\wi}(2, \OO) $ \\
\hline
$ C_2 \! \equiv \! B_2 $  & 
$ A_3 \! \equiv \! D_3 $  & 
$ D_4 $  &
$ D_6 $  \\
$ \fra{sy}_{\wi}(4, \R) $ & 
$ \fra{sy}_{\wi}(4, \C) $ & 
$ \fra{sy}_{\wi}(4, \Q) $ & 
$ \fra{sy}_{\wi}(4, \OO) $ \\
\hline
$ B_4 $  & 
$ D_5 \! \equiv \! E_5 $  & 
$ D_6 $  &
$ D_8 $  \\
$ \fra{my}_{\wi}(4, \R) $ & 
$ \fra{my}_{\wi}(4, \C) $ & 
$ \fra{my}_{\wi}(4, \Q) \qquad $ & 
$ \fra{my}_{\wi}(4, \OO) $ \\
\hline
\end{tabular}\hfill}

\medskip

In the case $\wi=1$, and through some of the low-dimension isomorphisms of
Lie algebras, this square coincides with the very nice form $\so(m,n), m=
2,3,5,9, n=0,1,2,4$ proposed by Sudbery \cite{Sud}. Only the first row is
different when $\wi =-1$, and a degenerate form correspond to $\wi =0$.


\section{Conclusions}

Consideration of the three $\sa, \sl, \sy$ CK series leads in a rather
natural way to a ``tower" of ``classical" \F-like squares relating 
different algebras in these CK families. In the lowest dimensional cases
$N =1, 2$, octonions are also allowed, and in these cases these
squares can be extended to $\mytimes 4$. When the constants $\w_i$  are
different from zero, we get simple Lie algebras in these three series and
the ``classical" $N=2$ \F\ square reduces to the  different versions
(compact and non-compact) of the original \F\ square, while $N=1$ gives the
one proposed by Sudbery. 

The introduction of an adequate notation is the key to open the view to
this square: in the standard conventional notation  and in the special case
where all constants $\w_i$ are different from zero, this square would read
in a rather uninspiring form whose direct relation to reals, complex
numbers and quaternions is rather remote:
\be
\begin{array}{lll}
\so(p,q)     & \su(p,q) & \sp(p,q) \\
\sl(\npo, \R) & \sl(\npo, \C) & \su^*(2(\npo)) \\
\sp(2(\npo), \R) & \su(\npo, \npo) & \so^*(4(\npo))  
\end{array}
\ee

This approach also displays the basic role played by the orthogonal CK
family, $\so_{\win}(\npo) \equiv \sa_{\win}(\npo, \R) $. This is the
antihermitian CK algebra over the reals, and appears as a subfamily in all
other CK series. To gain familiarity with all CK families, the orthogonal
one should be first studied at depth. 

We finally mention that similar ideas might also be pursued for the
remaining (orthogonal and symplectic) CK families, which are associate to
Lie algebras of ``antisymmetric" or ``symplectic-antisymmetric" 
(instead of antihermitian) matrices. Details for the case $\K=\Q$ are
involved. Three more towers (linear, orthogonal and symplectic) appear. 
Each tower makes sense for CK algebras with a fixed set of constant values
$\w_i$, and there are as many such towers as the
$3^N$ essentially different sets of constants $\w_i$. Different  real
forms, either compact or not, of {\em all} simple real Lie algebras are
related among themselves by several of these \F-like squares.


\section*{Acknowledgments}

M.S. would like to acknowledge A. Sudbery for his useful comments. 


\end{document}